\documentstyle[prl,aps,multicol]{revtex}
\input{epsf}
\begin{document}
\draft
\title{Fractal fluctuations in quantum integrable scattering}  
\author{
 Italo Guarneri$^{(a,b,c)}$, and Marcello Terraneo$^{(a,b)}$
}  
\address{
$^{(a)}$International Center for the Study of Dynamical 
Systems,
} 
\address{
Universit\`a degli Studi dell'Insubria, Via Valleggio 11,  
22100 Como, Italy} 
\address{$^{(b)}$Istituto Nazionale per la Fisica della Materia, 
Unit\`a di Como, Via Valleggio 11, 22100 Como,  Italy}      
\address{$^{(c)}$Istituto Nazionale di Fisica Nucleare, 
Sezione di Pavia, Via Bassi 6, 27100 Pavia, Italy}   
\date{\today}
\maketitle

\begin{abstract}
We theoretically and numerically demonstrate  that completely integrable 
scattering processes may 
exhibit fractal transmission fluctuations, due to typical 
spectral properties of integrable systems.
 Similar properties  also occur with scattering processes in 
the presence 
of strong dynamical localization, thus explaining recent numerical 
observations of fractality in the latter class of systems. 

\end{abstract}

\pacs{PACS numbers: 05.45.Df, 05.40.-a, 05.45.Mt}  

\begin{multicols}{2}
\narrowtext

\noindent
Scattering processes dominated by a statistically large number of 
metastable states display  reaction rates which depend 
on energy or other control parameters in  complicated ways. 
Reaction curves ({\it{i.e.}}, curves    
obtained by plotting cross sections versus the relevant parameters)  
 demand statistical description. This approach,   
 originated in nuclear physics, has given rise to 
the theory of quantum chaotic scattering\cite{fyod}. Classical chaotic 
dynamics plays an important role therein \cite{smila}, 
with applications in many 
fields, 
including  the study of mesoscopic conductance fluctuations\cite{jala}.\\
In the presence of complicated reaction curves, a natural question 
is about their 
fractality. The latter is not meant in the strict mathematical sense, 
but rather as a property to be possibly observed on a wide   
range of resolution scales.  
Reaction curves probe the location of resonance poles in the complex 
energy plane, so   
their fluctuation properties - including fractality - are encoded 
in the distribution of such poles. In particular, the lack of smoothness 
required by  fractality can only be produced by resonance poles clustering 
in the vicinity of the real energy axis. This excludes fractal reaction 
curves in quasi-classical cases when the underlying classical dynamics 
is  completely chaotic ({\it i.e}, uniformly hyperbolic), 
because the quantum resonances are then  
concentrated away from the real axis. This  reflects  the exponential 
decay in time exhibited by such systems over long time scales.
A dynamical signature of poles clustering near the real axis is 
instead slow algebraic decay of the survival probability inside the 
interaction region. 
 On the classical level, such slow  decay 
is in particular exhibited by  
systems with a mixed phase space, endowed with a hierarchical 
structure of stable islands\cite{geis}.
 On such grounds, fractal fluctuations 
were predicted by Ketzmerick for the quasi-classical 
transmission of electrons through 
mesoscopic cavities\cite{rol}.
This prediction has received numerical\cite{CGM} and experimental 
\cite{expe} support. However, fractal fluctuations have been 
numerically observed also in 2d tight-binding models of quantum dots, 
where the relation to classical dynamics is unclear\cite{spain}, and even 
in models where this relation  is irrelevant\cite{BCGT}, due to strong 
quantum  localization on one hand, and to absence 
of significant classical critical structures on the other.
It therefore appears that fractality of reaction curves is not strictly 
associated with critical structures in the classical phase 
space. 

The issue of fractality in the remaining major 
class of dynamical systems, namely the integrable ones, 
has not yet been investigated\cite{jala1}. This is the main purpose of this 
Letter. 
We first identify   
special statistical  properties of resonance 
poles, which afford fractal reaction curves. This we do in absolute 
generality, without assuming integrability or other special properties 
of the quantum system. We then 
argue  that 
such properties are optimally  exhibited  by suitable, completely integrable 
processes, thanks to a well known generic property 
of integrable systems, that their energy levels share some of the properties 
of a random sequence\cite{berry}. 
We use a textbook scattering problem to confirm 
our theory with numerical data. Finally  we note that the  same 
properties are also typical of fully chaotic, yet strongly localized 
systems, and thus we explain  the results of ref.\cite{BCGT}.

We consider a weakly open quantum system, with scattering resonances at 
$E_j-i\Gamma_j/2$ and the real energies $E_j$ arranged in increasing 
order. 
The energy dependence of a typical cross section $T(E)$ consists of 
a smooth background plus a resonant part, which we write in the form:
\begin{equation}
\label{reso}
T_r(E)=\sum_j c_j\frac{\Gamma_j^2}{(E_j-E)^2+\Gamma_j^2}
\end{equation}
with $c_j$ slowly varying with $j$. We restrict within an energy interval 
$(E_0-W,E_0+W)$, and we assume  $\epsilon \ll W \ll E_0$, where $\epsilon$
is the 
mean 
level spacing.
To perform 
fractal analysis of the graph of $T_r(E)$ vs $E$ 
we divide the interval $(E_0-W,E_0+W)$ into subintervals $\Delta_k$, 
$(k=1,2,...,M)$ 
of equal size $\delta\propto 1/M$. Upon each subinterval we pile up squares
 of side $\delta$, and  denote by $N(\delta)$ 
the total number of squares met by  
 the graph of $T_r(E)$ . Algebraic scaling 
$N(\delta)\propto\delta^{-f}$ with $f>1$ between scales $\delta_{min} \ll
\delta_{max}$ signals that in between such scales the graph exhibits  
a fractal (box-counting) dimension $f$. In order to determine  $f$ we compute
$$
\frac{\log (N(\delta))}{\log(\delta^{-1})}\approx
2+\frac{\log(M^{-1}\sum_k\sigma_k(T_r))}{\log (M)}
\approx 2-\frac{\log\langle \sigma(T_r)\rangle}{\log (\delta)}
$$
where $\sigma_k(T_r)\geq 0$ is the maximal excursion of $T_r(E)$ within 
the $k-$th interval.
We next assume the following. First, in the given energy interval, 
  the frequency $P(\Gamma)$  
of widths less than $\Gamma$ scales like $P(\Gamma)\sim a\Gamma ^{1-\alpha}$ at 
$\epsilon<\Gamma<{\overline\Gamma}$, with ${\overline \Gamma}$ the 
mean resonance width, and $0<\alpha<1$.  
Second, both  $\Gamma_j$ 
and  $E_j$ form 
uncorrelated sequences. Finally, resonances are strongly overlapped, 
in the sense that $\epsilon
\ll {\overline\Gamma} \ll W$. Then, at 
$\epsilon \ll  \delta \ll {\overline\Gamma}$, the fluctuation of $T_r$ in
an 
interval $\Delta_k$  is 
mostly due to many tiny resonant peaks which are centered inside the interval 
and are narrower than $\delta$. Every such peak contributes a Lorentzian term 
in (\ref{reso}), and 
the mean square  oscillation of this  term as  
 $E$ ranges in $\Delta_k$ is $\sim\delta^{-1}P(\delta)^{-1}
\int_0^{\delta}dP(\Gamma)\Gamma\sim 1$.
There are $n(\delta)\sim \epsilon^{-1}\delta P(\delta)$ such peaks; as 
they contribute  uncorrelated oscillations, we estimate 
$\langle\sigma(T_r)\rangle\sim \sqrt{n(\delta)}\sim
\epsilon^{-1/2}\delta^{1-\alpha/2}$. Therefore, 
$f=1+\alpha/2$ in a range $\delta_{min}<\delta<\delta_{max}$, with 
$\delta_{max}\sim{\overline\Gamma}$, and $\delta_{min}$ 
roughly estimated by $n(\delta_{min})\sim 1$, that is, 
$\delta_{min}\sim (\epsilon/a)^{1/(2-\alpha)}$.

Fluctuating cross sections may also be generated at fixed energy, by 
varying other parameters, as in the case of magnetoresistance fluctuations 
in mesoscopic physics. The above 
analysis carries over to such fluctuations on replacing energy by the 
relevant parameter, provided that the above assumptions 
remain satisfied. 

The above described conditions  are  met in some physically
relevant 
situations.  
Uncorrelated sequences of energy levels are  a distinctive feature
of generic integrable systems\cite{berry}. We then surmise fractal reaction 
curves 
for completely integrable scattering processes, provided they 
 display 
a slow decay, leading to an inverse power law distribution  of $\Gamma$'s.  
 We shall presently describe an explicit example of such a process.

We consider the quantum dynamics of a particle of unit mass moving 
inside   the infinite strip $0\leq x\leq L_x$ in the $(x,y)$ plane, with 
 tunneling barriers 
 at $y=\pm L_y/2$. The Hamiltonian is
$$
H=-\frac{\hbar^2}{2}\Delta + \hbar\sigma\delta(y-L_y/2)+
\hbar\sigma\delta(y+L_y/2)
$$
We use periodic boundary conditions at $x=0,\;x=L_x$. The Dirac delta 
functions 
enforce additional boundary conditions: $\partial_y\psi(x,y+)-
\partial_y\psi(x,y-)=2\sigma\hbar^{-1}\psi(x,y)$ at $y=\pm L_y/2$ .  The physical 
model is a rectangular billiard, 
whence the particle can escape 
into semi-infinite leads, by tunneling through the two 
horizontal sides. 
For $|y|>L_y/2$ eigenfunctions are  
$u_{E,m}^{\pm}(x,y)=\phi_m(x)\theta_k^{\pm}(y)$, $\phi_m(x)=\exp(2\pi i mx/L_x)$, 
$\theta_k^{\pm}(y)=
A^{\pm}(k)\exp(ik|y|)+B^{\pm}(k)\exp(-ik|y|)$ 
 and 
$2E/\hbar^2=k^2+4\pi^2m^2/L_x^{2}$, with $\pm$ denoting the upper 
($y>L_y$) and the 
lower ($y<-L_y$) lead respectively. 
For given $E>0$ there are a finite 
number 
of open scattering channels labelled by 
the integer $m$, $|m|\leq {\rm Int}(L_x(\pi\hbar)^{-1}\sqrt{E/2})$,  
 and by the lead label 
$\pm$. 
The coefficients $A^{\pm}(k)$ are related to $
B^{\pm}(k)$
by the scattering matrix. 
Scattering resonances are located at complex values of energy $
z_{n,m}=E_{n,m}-i\Gamma_{n}/2
=2\pi^2m^2\hbar^2/L_x^{2}+
k^2_n\hbar^2/2$, ($n=1,2,...$), where $k_n$ are the  complex roots of the 
equations: 
\begin{equation}
\label{trasc}
e^{ikL_y}\pm 1 \mp ik\hbar\sigma^{-1}=0
\end{equation}
 The following asymptotic formulae can be computed by using Lagrange's 
theorem on the inversion of analytic functions:
\begin{equation}
\label{gamma}
\begin{array}{c}
E_{n,m} 
= 
E^{0}_{n,m}-
\frac{\hbar v_n^2}{L_y\sigma}\left\{
\frac{\sigma}{v_n}\arctan 
(\frac{v_n}{\sigma})+O\left(\frac{\hbar}{\sigma L_y}\right)
\right\}\nonumber\\

\Gamma_n 
= 
\frac{\hbar v_n^3}{L_y\sigma^2}\left\{
\frac{\sigma^2}{v_n^2}\log (1+\frac{v_n^2}{\sigma^2})+
O\left(\frac{\hbar}{\sigma L_y}\right)\right\}\\
\end{array}
\end{equation}
where  $v_n=n\pi\hbar/L_y$ is the velocity in the 
$n-$th vertical mode of the closed rectangle, and 
 $E^{(0)}_{n,m}=
2\hbar^2\pi^2m^2/L_x^{2}+
v_n^2/2$ are the eigenvalues of the closed $(\sigma=\infty)$
billiard. 
The 1st term on the rhs of the 2nd eq.(\ref{gamma}) is the decay rate 
of a classical billiard ball inside the closed rectangle, with  
velocity $v_n$ 
in the $y-$direction, and  absorption probability at $y=\pm L_y/2$ equal to 
the transmission coefficient for a plane wave through   
 on a $\delta$ barrier, given by $v_n^2(\sigma^2+v_n^2)^{-1}$. 
 We assume $\hbar \ll \sigma L_y$ and thereby neglect $O(\hbar/\sigma
L_y)$ 
corrections in (\ref{gamma}). The statistics of real parts of 
resonances is then like 
the energy level statistics of an integrable system, with a mean level 
spacing only different  by corrections of order $(\hbar/\sigma L_y)^3 $ 
from that 
of the closed billiard: $\epsilon=2\pi\hbar^2(L_x L_y)^{-1}$. 
The correction to the closed billiard levels in (\ref{gamma}) lifts 
possible degeneracies due to commensurate geometry. 

In an energy interval $(E_0-W,E_0+W)$, $\epsilon\ll W \ll E_0$,
the smooth (Thomas-Fermi) part of the integrated 
$P(\Gamma)$ distribution 
 is computed 
from a microcanonical distribution of classical billiard trajectories 
at energy $E_0$, each with a decay rate $\Gamma$ (\ref{gamma}).  
This quasi- classical distribution has a mean 
${\overline\Gamma}_{E_0}\simeq \hbar\sigma^{-2}L_y^{-1}E^{3/2}_0$, 
and behaves like $
a\Gamma^{1/3}$ at small $\Gamma$, with $a=(\pi^2E_0/2)^{-1/2}
(L_y\sigma^2/\hbar)^{1/3}$. 
 The same behaviour can be assumed 
for the quantal distribution, provided that 
$E_0\leq\sigma^2$, and that $\epsilon \ll {\overline\Gamma}_{E_0}$. 
The former condition 
ensures $\Gamma^{1/3}$ behaviour of the 
quasi-classical distribution in a 
range of $\Gamma-$values comparable to ${\overline\Gamma}_E$. 
Together with the  latter 
condition  
it  ensures that the 
quasi-classical behaviour is observed 
over a statistically significant number of resonances.
Reflecting  generic properties of energy spectra of  
integrable systems,  the real 
parts of resonances, arranged in increasing order, form an 
essentially uncorrelated sequence. 
At fixed quantum number $m$, they form an ordered 
ladder, but the superposition of a large number of different, 
uncorrelated  
ladders results in a Poisson-like statistics. On the same grounds  
we assume uncorrelated  resonance widths $\Gamma$, too.

The transmission amplitude at energy $E$ from the lower $m-$ channel to the 
upper $l-$ channel is:  
\begin{equation}
\label{ampl}
S_{m-,l+}(E)=\frac{k^2(E,m)\hbar^2}{\sigma^2 e^{2ik(E,m)L_y}+
(k(E,m)\hbar+i\sigma)^2}\delta_{ml},
\end{equation}
where $k(E,m)=\sqrt{2E\hbar^{-2}-4\pi^2 m^2 L_x^{-2}}$.
A computation shows that the residue of (\ref{ampl}) at a resonance pole 
 is $\sim 
-i \Gamma /2$ at small $\Gamma$. 
 Hence, the resonant part 
of the total transmission 
coefficient:
\begin{equation}
\label{trasm}
T(E)=\sum_m |S_{m-,m+}(E)|^2
\end{equation}
(the sum being over all open channels at energy $E$) has the form 
(\ref{reso}) apart from a slowly varying factor.
 The smooth part is $\propto E^{5/2}/\sigma^4$.  
Collecting various estimates we see that at  
$\hbar/(\sigma L_y) \ll 1$, $\hbar^2/L_y^2 \ll W \ll E_0
<\sigma^2$, $E^{3/2}_0L_x\hbar^{-1}\sigma^{-2}\gg 1$
the assumptions  of our general argument are satisfied, hence 
 the graph of $T(E)$ vs $E$ in $(E_0-W<E<E_0+W)$ should be fractal   
with dimension $f=4/3$, over scales intermediate between $
\delta_{min}\sim 1.8\epsilon^{3/4}{\overline\Gamma}_{E_0}^{1/4}$ and 
$\delta_{max}\sim {\overline\Gamma}_{E_0}$.

A numerically computed graph of $T(E)$ is shown in Fig.1 (lower).  
The corresponding 
fractal analysis is shown in Fig.2 and fully confirms the theory. 
In that case, the above estimates for  $\delta_{min}$, $\delta_{max}$ give 
$10^{1.9}$ and $ 10^{3.9}$ respectively.

A generalization of the above model allows 
for the investigation of parametric fluctuations. It is obtained  
by using  boundary 
conditions $\psi(0,y)=e^{i\phi}\psi(L_x,y)$. This is equivalent 
to a particle moving on a cylinder, with axis in the $y-$ 
direction,  enclosing a magnetic flux $\phi$. 
This problem is still completely integrable. 
Replacing $m$ by $m-\phi/(2\pi)$ throughout the equations derived at 
$\phi=0$ yields the corresponding theory. At fixed energy $E$, the total 
transmission fluctuates as $\phi$ is varied, in the manner illustrated in 
Fig.1 (upper). The theory of such fluctuations is completely parallel to the 
one we have described for fluctuations vs energy at fixed $\phi=0$. 
Resonances depend on $\phi$,  $z_{n,m}=z_{n,m}(\phi)$, and the complex 
values of $\phi$ solving the equations   $z_{n,m}(\phi)=E$ define 
resonance poles in the complex $\phi$-plane. Omitting computational details, 
 the real parts of such poles are distributed 
in $[0,2\pi]$ with a mean spacing 
$\epsilon_{\phi}\sim \pi^2\hbar\sqrt{2/E}/L_x$. 
The distribution of their widths $\Gamma_{\phi}$ behaves 
like $\Gamma_{\phi}^{1/3}$  
at small $\Gamma_{\phi}$, with a mean $\sim L^2_x{\overline\Gamma}_E
\epsilon_{\phi}/(8\pi^2\hbar^2)$. Our general discussion is thus valid 
for parametric fluctuations, too. In fact 
 numerical data shown in Fig.3 
 demonstrate the predicted fractal dimension $4/3$.
\begin{figure}
\centerline{\epsfxsize=6.6cm\epsffile{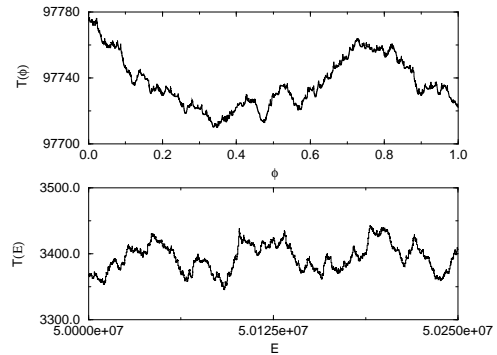}}
\caption{
Total transmission versus flux $\phi$ at fixed energy 
$E=5 \times 10^{10}$ (upper plot), and versus energy $E$ at fixed flux
$\phi=0$ 
(lower plot), for $\sigma=3.5\times 10^5$ and $\sigma=10^4$ 
respectively. In both cases $\hbar=1$, $L_x=2$, $L_y=0.4$. The energy
range in  the lower plot is $(E_0,E_0+W)$ with $E_0=5 \times
10^7$, $W=2.5 \times 10^5$. Units are 
arbitrary.}
\label{fig1}
\end{figure}

\begin{figure}
\centerline{\epsfxsize=6.7cm\epsffile{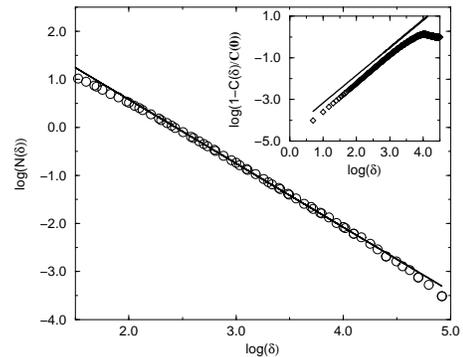}}
\caption{Fractal analysis of the graph of $T(E)$ vs $E$ 
shown  
in the lower fig \ref{fig1}. The meaning of $\delta, N(\delta)$ is
explained in the 
text. 
 The straight line 
corresponds to fractal dimension  $f=4/3$. 
The inset shows the autocorrelation $1 - C(\delta)/C(0)$ vs $\delta$. 
The straight line has slope $4/3$.
  }
\label{fig2}
\end{figure}

Both for energy-dependent and for parametric fluctuations 
 we have computed autocorrelations of the fluctuation graphs. 
For the case of fluctuations vs energy, such correlations are defined by
$$
C(\delta)=\langle T^{(fl)}(E)T^{(fl)}(E+\delta)\rangle_E
$$
where $T^{(fl)}$ is obtained from $T$ by subtracting 
a smooth, slowly varying part, and  
the average is taken  over the scanned interval of energies. 
Treating  $T^{(fl)}(E)$ as a stationary stochastic process, one easily 
finds that such correlations behave at small 
$\delta$ like $C(0)-{\rm const.}\times\delta^{2\gamma}$, where 
$\gamma$ is the scaling exponent of the rms increment 
of $T(E)$ over intervals of length $\delta$: 
$\langle |T(E+\delta)-T(E)|^2\rangle_E \propto\delta^{2\gamma}$ (note
that 
 subtracting the smooth part does not alter fractional scaling).  
Generally speaking, the rms {\it increment} is a quite different 
 quantity from the average {\it excursion} 
 which enters the definition of the fractal 
dimension. With the present strong statistical properties the two 
quantities scale in the same way at small $\delta$: $2 \gamma = 2
-\alpha$, so 
$C(\delta)\sim C(0)-{\rm const.}\times \delta^{4/3}$, as confirmed by 
numerical data. However fractional scaling of correlations 
is not in general a sufficient condition for fractality. For instance, 
 correlations in Fig.2 and in  Fig.3 exhibit a $4/3$ scaling down to small 
$\delta$ scales below the fractal range. This is because they 
are still determined by 
the statistics  of narrow, non-overlapped individual peaks, which do not 
produce  fractality any more. 
On the other hand,  on increasing 
$\delta$  
correlation functions depart  from the predicted fractional behaviour 
already at values well  within the fractal range, because 
 a larger statistics is needed for {\it increments} recorded 
over a finite $\delta-$grid to sample the distribution of 
{\it excursions} over the same grid. 
\begin{figure}
\centerline{\epsfxsize=6.7cm\epsffile{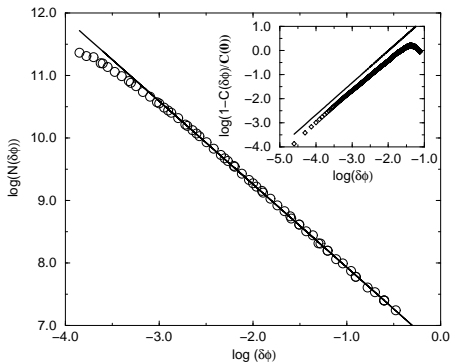}}
\caption{Fractal analysis of the flux-dependent
fluctuations shown in the upper fig.1.  
The straight line corresponds to $f=4/3$ . The inset shows the 
correlation scaling exponent  $2\gamma=4/3$.}
\label{fig3}
\end{figure}
The work reported in this  Letter hinges on the fully general fact, 
that fractality of scattering 
fluctuations is, first and foremost, a matter of complex level statistics.
As in the case of real level statistics (closed systems), integrable 
systems have special properties in this respect, and 
we have in fact demonstrated fractal reaction curves for a completely 
integrable 
process. A few remarks are in order about the relevance of this 
finding:

\noindent
(1) In completely integrable scattering processes there is no mixing 
of flux between different channels. Like  pseudo-randomness of energy 
levels, fractality   
comes of superimposing non-fractal fluctuation patterns 
from different, uncorrelated channels. Therefore the nature of the reaction 
curves depends on the relative weight assigned to different channels.
 
\noindent
(2) The present result may also be  relevant to quasi-integrable systems, 
which  possess large stable components in their phase space. 
Such components  may contribute a significant set of resonances, produced 
by tunneling through invariant manifolds, with the statistical properties 
considered in this Letter. Their interplay with  critical structures
at the border of the stable regions demands careful analysis.

\noindent
(3) uncorrelated energy spectra also 
occur with fully chaotic systems in the regime of strong quantum 
localization. A prototype system in this class is the kicked rotor, 
the quasi-energy spectrum of which has a Poisson-like statistics  
in case of strong localization\cite{pois}. For this class of systems, 
the $\Gamma$ (differential) distribution behaves like $1/\Gamma$ down to very 
small scales\cite{CMS,TG}. 
This explains recent findings\cite{BCGT} of {\it parametric} fractal 
fluctuations with dimension $3/2$, detected in the {\it survival probability} 
at fixed time, on varying a magnetic flux $\phi$. 

Support from MURST Research Project ``Chaos and localization in classical and
quantum mechanics'' is gratefully acknowledged.

\end{multicols}


\begin{references}
\bibitem{jala1} Transport fluctuations in 
 integrable cavities have been studied 
in detail, but their fractality doesn't appear to have been investigated  
as yet. See, e.g.,  ref.\cite{jala}.
\end{references}
\end{document}